\documentclass[twocolumn,amsmath,amssymb,floatfix]{revtex4}

\usepackage{CJK}
\usepackage{csquotes} 
\usepackage{graphicx}
\usepackage{dcolumn}
\usepackage{color}
\usepackage{amsmath}

\usepackage{hyperref}

\renewcommand{\vec}[1]{\boldsymbol{\mathbf{\mathrm{#1}}}}

\begin{document}

\begin{CJK*}{GB}{gbsn}

  \title{Triaxiality explored by an odd quasi-particle}

  \author{Weichuan Li~(\CJKchar{"C0}{"EE}\CJKchar{"CE}{"B3}\CJKchar{"B4}{"A8})}
  \author{Stefan Frauendorf}
  \email{sfrauend@nd.edu}
  \author{Mark A.~Caprio}
  \affiliation{Department of Physics and Astronomy, University of Notre Dame, Notre Dame, Indiana 46556-5670, USA}

  \date{\today}

  \begin{abstract}
    The triaxiality of odd-mass nuclei is investigated by coupling a
    quasiparticle to an even-even core through the core-quasiparticle coupling
    model. Both soft and rigid triaxial cores are considered. The \enquote{soft
      core} is described by the collective model with rotation-vibrational
    motion, while the \enquote{rigid core} is described by the triaxial rotor
    model, which is a limiting case of the collective model with only rotational
    motion.  We show that the presence of the odd quasiparticle modifies the
    collective quadrupole dynamics of the core to appear more \enquote{rigid}.
  \end{abstract}

  \maketitle

\end{CJK*}

\section{Introduction}\label{sec:intro}

About 85 percent of the known atomic nuclei are proposed to have an intrinsic
axial prolate shape, while triaxial shapes have been predicted to appear in the
mass regions around $A=80$, 100, 130, and 190 (see, e.g., Ref.~\cite{Moller}).
Evidence for strong deviation from axiality in even-even nuclei has been
discussed in the framework of the Interacting Boson Model
\cite{stagger,staggeringOddandEven} and collective Bohr Hamiltonian
\cite{collectiveH}.  The structure of the quasi-$\gamma$-band provides the best
available information, which indicates the prevalence a soft mode. As recently
reviewed by Ref.~\cite{background}, the staggering between the even-$I$ and
odd-$I$ members seems to indicate that the majority of triaxial even-even nuclei
are soft, and only a few triaxial even-even nuclei show a certain
$\gamma$-rigidity.
 
On the other hand, it has been known for a long time that the behavior of
odd-$A$ nuclei can be well described by the quasiparticle triaxial rotor (QTR)
model pioneered by Refs.~\cite{Meyer-ter-Veen,TokiFaessler} and used numerous
times later on. More recently the QTR provided the basis for analyzing phenomena
arising from triaxiality in the presence of one or several quasiparticles.  One
phenomenon is wobbling motion, which was predicted to appear in even-even
nuclei~\cite{BohrMottlelsonII}. However it was first found experimentally in the
odd-$A$ nuclei $^{163-167}$Lu at high spin~\cite{Luexperiment} and analyzed in
the QTR framework \cite{Hamamoto}. Extending this QTR analysis,
Ref.~\cite{transverse} introduced the concepts of transverse and longitudinal
wobbling. The authors classified the Lu isotopes as tranverse and predicted
$^{135}$Pr as another example of transverse wobbling, which was experimentally
confirmed by Ref.~\cite{pr135}.
  
The second phenomenon is the possibility of a chiral geometry when two or more
quasiparticles are coupled to a rigid triaxial rotor \cite{FRMeng}.  Evidence
for chiral doubling has been found in all mass regions where triaxiality is
expected (see review \cite{chiralReviews}). This raises the question why the
wobbling and chiral bands are well accounted for by quasiparticles coupled to a
rigid triaxial rotor while the quasi-$\gamma$-bands in the even-even-neighbors
point to a soft triaxial core.
  
The present paper addresses the apparent conflict between the evidence for a
soft triaxial mode in even-even nuclei and a rigid triaxial core in their
odd-$A$ neighbors.  Quasiparticles coupled to a triaxial rotor constitute a well
developed model. Coupling quasiparticles to a soft core is not so well
explored. In this paper we use the core-quasiparticle coupling (CQPC)
model~\cite{cqpc}, which describes the coupling of one quasiparticle with the
most general collective quadrupole excitations of the core.  A \enquote{soft
core} is described by the collective model with rotation-vibrational motion,
which encompasses the \enquote{rigid core} triaxial rotor model as a limiting
case of the collective model with only rotational motion.  We show that the
presence of the odd quasiparticle modifies the collective quadrupole dynamics of
the core in a major way, making it appear more rigid.
    
So far, there is systematic experimental evidence for the wobbling motion mode
in the odd-mass triaxial nuclei of the A=130 and 160 regions.  In this paper,
results for A$\approx$110 nuclei are presented. The systematic appearance of the wobbling mode
in the odd-mass Pd, Ru and Rh isotopes is predicted. The predictions are
confirmed by a recent experiment on $^{105}$Pd \cite{105Pd}.
   
Section \ref{sec:models} describes the models for the soft and rigid cores, as
well as the input parameters. Section \ref{sec:discussion} presents the
numerical results for the Pd, Ru and Rh isotopes and discusses their
implications.  These results were reported in part in Ref.~\cite{thesisLi}.
  
\section{Theoretical background}\label{sec:models}

In this paper we study the coupling of an odd quasiparticle to two types of even-even cores, which we distinguish as ``soft'' and ``rigid''. 

\subsection{Soft core}

The collective model is used to describe the soft even-even core nucleus. The
nuclear shape is described in terms of a nuclear surface deformation.  The
radius of the nuclear surface $R(\theta,\phi)$ is described as
follows:
\begin{equation}
R(\theta ,\phi )=R_{0}(1+\sum_{\mu }^{\ }\alpha _{2 \mu }^{*}Y_{2 \mu }(\theta ,\phi )), 
\end{equation}
where $\alpha _{2 \mu }$ are collective coordinates. In the body fixed frame, the
collective coordinates are expressed in terms of $a_{0}$, $a_{2}$ and the Euler
angles $(\Omega=(\theta_{1},\theta_{2},\theta_{3}))$, where $a_{0}=\beta \cos
\gamma $ and $a_{2}=\frac{\beta}{\sqrt{2}}\sin\gamma$, with $\beta$ being the
deformation parameter and $\gamma$ the asymmetry
parameter~\cite{BohrMottlelsonII}.
 
The collective Bohr Hamiltonian is written in terms of angles $\theta_{i}$ of rotation about the three principal axes of the shape and the shape coordinates $\beta$ and $\gamma$ as
\begin{equation}
  \label{eq:H}
H = T+V.
\end{equation}
The  kinetic energy operator has the standard irrotational flow form 
\begin{multline}
T=-\frac{\hbar^{2}}{2B}\left[\frac{1}{\beta ^{4}}\frac{\partial }{\partial \beta }
  {\beta }^{4}\frac{\partial }{\partial \beta }
  +\frac{1}{\beta ^{2}\sin 3\gamma }\frac{\partial }{\partial \gamma }\sin 3\gamma \frac{\partial }{\partial \gamma }\right]
\\
+H_{\mathrm{rot}},
\end{multline}
where 
\begin{equation}\label{Hrot}
H_{\mathrm{rot}}=\sum_{i=1}^{3}\frac{R_{i}^{2}}{2\mathcal{J}^{}_{i}}.
\end{equation}
describes the rotational energy of the nucleus. Here, 
 $R_{k}=\frac{\hbar}{i}\frac{\partial}{\partial\theta_{i}}$ represent the angular momentum operators in the body-fixed frame and 
 \begin{equation}\label{MoI}
 \mathcal{J}_{k}=4B\beta^{2}\sin (\gamma -\frac{2}{3}\pi k)^{2}
 \end{equation}
  the moments of inertia. 
 
\begin{figure}[t]
  \begin{center}
   \includegraphics[width=\hsize]{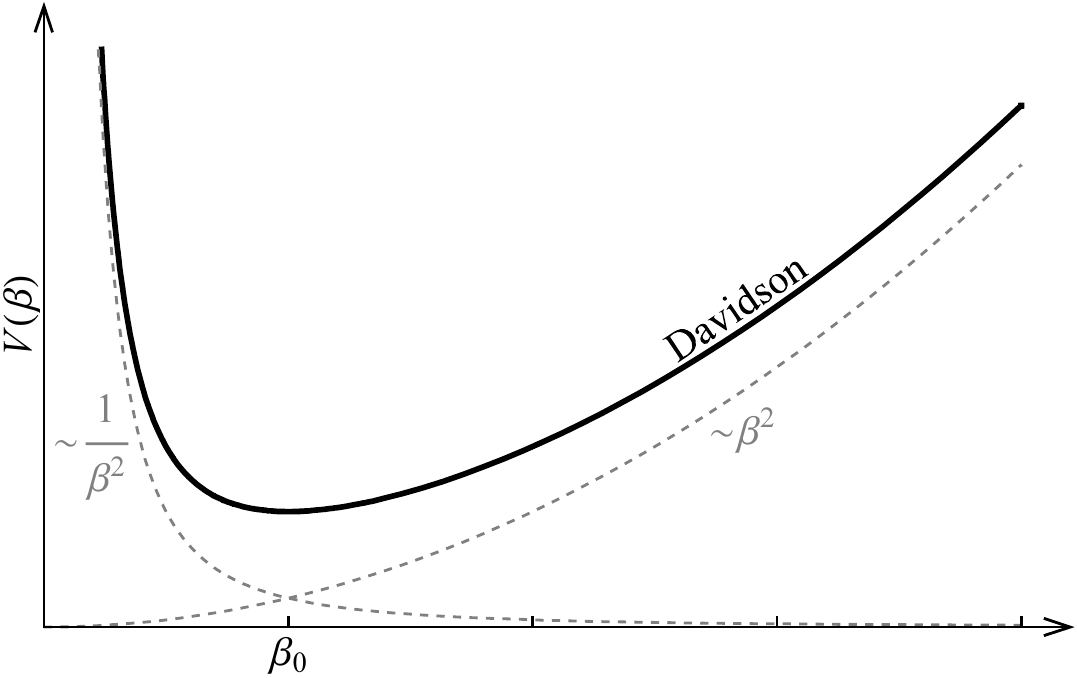}
  \end{center}
     \caption{The $\beta$ dependence of the potential $V(\beta,\gamma)$ in~(\ref{eq:davidson}), given by a Davidson potential in $\beta$. The minimum of the potential is at $\beta_{0}$.  The asymptotic dependences $\sim\beta^{-2}$ and $\sim\beta^{2}$ are indicated by dashed lines.
       \label{fig:ddavidson}
       }
\end{figure}

\begin{figure}[t]
  \begin{center}
  \includegraphics[width=\hsize]{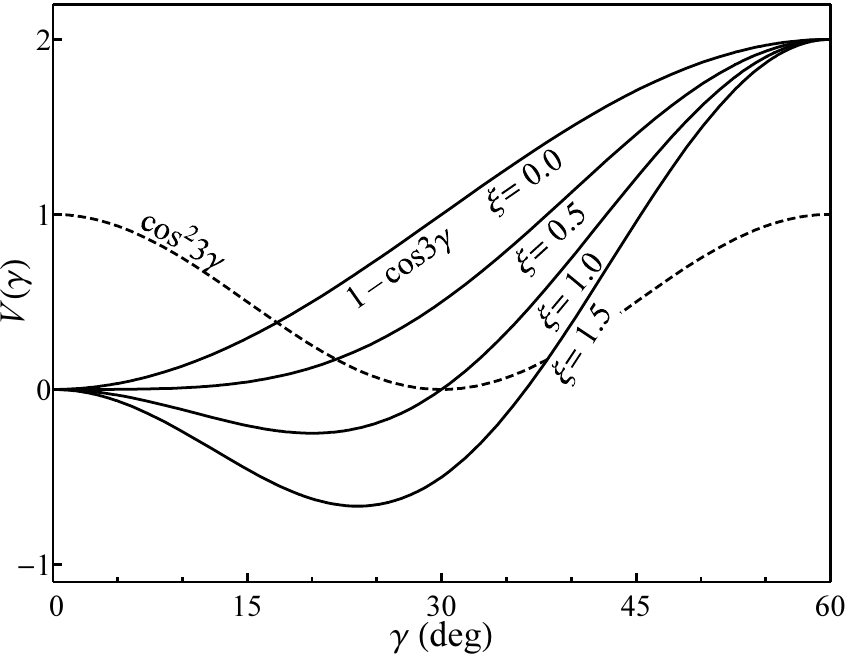}
  \end{center}
  \caption{$\gamma$ potential V($\gamma$).
The $\gamma$ dependence of the  potential $V(\beta,\gamma)$ in~(\ref{eq:davidson}), for
  various $\xi$, showing the dependence of the position of the minimum on $\xi$, for fixed $\chi=1$ (solid lines).  The shape of the
  contribution from  $\cos^2 3\gamma$ is also shown (dotted line).
  Figure from Ref.~\cite{collectiveH}.
  \label{fig:gamma}
  }
\end{figure}

For the potential $V(\beta ,\gamma )$ as a function of deformation, we use a
triaxial Davidson (TD) potential, that is, a Davidson potential with respect to
$\beta$~\cite{elliott1986:gsoft-davidson}, combined with a 
potential that is soft with respect to the triaxiality parameter $\gamma$~\cite{iachello2003:y5}:
\begin{multline}
  \label{eq:davidson}
  \frac{V(\beta ,\gamma )}{V_0}
  =\beta_0^2\left(\frac{\beta_{0}}{\beta}-\frac{\beta}{\beta_0}\right)
  \\
+\chi \left[(1-\cos 3\gamma )+\xi \cos 3\gamma ^{2}\right].
\end{multline}
It is readily apparent that the parameter $\beta_{0}$ in the first term
of~(\ref{eq:davidson}) serves as a scale parameter on $\beta$, and, in
particular, it defines the position of the minimum in $\beta$, as shown in
Fig.~\ref{fig:ddavidson}.  The parameter $\xi$ in the second term controls the
location of the minimum in $\gamma$~\cite{collectiveH,Caprio}, as illustrated in
Fig.~\ref{fig:gamma}, while $\chi$ controls the depth (or stiffness) of this
minimum in $\gamma$.

The collective Hamiltonian as formulated in~(\ref{eq:H})--(\ref{eq:davidson})
has five parameter degrees of freedom $(B,V_0,\beta_0,\chi,\xi)$.  However, two
of these degrees of freedom may be eliminated by use of analytic scaling
relations~\cite{caprio2003:gcm}: (1)~A simultaneous change in $B$ and $V_0$ of the
form $V_0\rightarrow aV_0$ and $B\rightarrow a^{-1} B$, simply multiplies the
Hamiltonian by an overall factor.  This rescales all eigenvalues ($E\rightarrow aE$),
while preserving ratios of (excitation) energies.  The wave functions and thus
electromagnetic matrix elements are left unchanged.  (2)~A simultaneous change
in $V_0$ and $\beta_0$ of the form $\beta_0\rightarrow a^{-1} \beta_0$ and
$V_0\rightarrow a^4V_0$ is recognized to yield a dilation transformation of the type defined
in~(5) of Ref.~\cite{caprio2003:gcm}.  This again preserves ratios of energies, but
dilates all wave functions in $\beta$, thereby introducing an overall scale factor ($\propto
\beta_0^2$) to all $E2$ matrix elements.  The energy and
$\beta$ scales may be chosen so as to obtain any desired values for the
observables $E(2^+)$ and $\langle 2 \Vert Q \Vert 0 \rangle$.

We are thus left with three remainining parameter degrees of freedom, which may
be taken as the potential parameters $(\beta_{0},\chi,\xi)$
in~(\ref{eq:davidson}).  These must be explored through numerical
diagonaliztion.  In this work, the Schr\"odigner equation is solved using the
algebraic collective model (ACM)~\cite{ACM,ACM2}, an algebraic framework for
numerically solving the collective model eigenproblem in an
$\mathrm{SU}(1,1)\times\mathrm{SO}(5)$ basis.
The energies of the lowest collective quadrupole excitations of the adjacent
even-even nucleus are used to determine these remaining parameters in the
collective Hamiltonian for the core.

In particular, the collective model parameters are chosen through a
least-squares fitting procedure so as to reproduce the ratios
$E(2_{2}^{+})/E(2_{1}^{+})$, $E(4_{1}^{+})/E(2_{1}^{+})$, and
$E(0_{2}^{+})/E(2_{1}^{+})$, together with the staggering of the quasi-$\gamma$ band, as
measured by the parameter $S(R)$ defined below in~(\ref{eq:S}). The staggering of the quasi-$\gamma$ band is the clearest signature to distinguish
between a $\gamma$-soft or a $\gamma$-rigid nuclear shape. For example
$S(4)=1.67$ for a $\gamma$ rigid triaxial rotor, and $S(4)=-2$ for a $\gamma$
independent potential~\cite{stagger}. The ratios $E(2_{2}^{+})/E(2_{1}^{+})$,
$E(4_{1}^{+})/E(2_{1}^{+})$, $E(0_{2}^{+})/E(2_{1}^{+})$~\cite{Casten}
sensitively depend on all three parameters the collective potential
(\ref{eq:davidson}) (see Ref.~\cite{collectiveH}). Through manipulating the
weights of the input data of the fit, parameter sets are found which in our view
best account for the experimental spectra. They are listed in
Table~\ref{tab:soft-params}.  The ACM solutions with these parameters provide
the relative excitation energies $E(R)/E(2^+)$ and the relative reduced matrix
elements of the quadrupole operator $ \langle R\Vert Q \Vert R' \rangle/Q_0 $,
which are inputs to the CQPC model. The scales $E(2^+)$ and $Q_0$ are treated as
free parameters, the choice of which is discussed below in
Sec.~\ref{sec:background:cqpc}.

\begin{table*}
\caption{\label{tab:soft-params} Parameters of the soft-core calculations, where  $\left( \beta_{0},\chi,\xi \right )$ define  the Davidson potential, $\kappa$ is the coupling strength of the quadrupole interaction, $E_{2}$ is the energy of the first $ 2^{+}$ state of the cores, and $\varepsilon_{j}-\lambda$ is the Fermi level relative to the neutron $h_{11/2}$ and the proton $g_{9/2}$  levels.
}
\begin{ruledtabular}
\begin{tabular}{ccccccc}
 Nucleus & $\beta_{0}$ & $\chi$ & $\xi$ & $\kappa$ & $E_{2}$ & $\lambda-\varepsilon_{j}$ \\ \hline
 $^{101}$Pd& 1.5 & 1.4 & 1 & 8.6 & 0.556 & -3.487 \\
  $^{103}$Pd & 1.8 & 1.6 & 2.6& 9.11 & 0.556 & -2.96\\
  $^{105}$Pd& 1.8& 2.2 & 2.8 & 9.92 & 0.512 & -2.47\\
  $^{107}$Pd& 2.1  & 0.6 & 3& 10.46 & 0.434 & -2.028\\
  $^{109}$Pd & 2.1 & 0.7 & 4 & 11 & 0.47 & -1.58 \\%
   $^{111}$Pd& 2.6& 0.5 & 3.1 & 9.36 & 0.425 & -1.0482 \\
    $^{113}$Pd & 2.7  & 0.9 & 3.6& 7 & 0.419 & -0.62 \\
     $^{115}$Pd& 2.4 & 0.9 & 3.8 & 8.7 & 0.426 & -0.184 \\
      $^{101}$Rh & 1.5 & 1.4 & 1.1 & 8.6 & 0.556 & 0.61  \\
       $^{103}$Rh& 1.8 & 1.6 & 2.6 & 9.11 & 0.556 & 0.63  \\
        $^{105}$Rh & 1.8 & 2.2 & 2.8 & 9.92 & 0.512 & 0.716  \\
         $^{107}$Rh& 2.1 & 0.6 & 3 & 10.46 & 0.434 & 0.76 \\
          $^{109}$Rh & 2.1  & 0.7 & 4 & 11 & 0.47 & 0.847  \\
           $^{111}$Rh& 2.6 & 0.5 & 3.1 & 9.36 & 0.48 & 0.697 \\
            $^{113}$Rh & 2.7  & 0.9 & 3.6  & 7 & 0.459 & 0.492 \\
             $^{115}$Rh& 2.4 & 0.9 & 3.8 &  8.7 & 0.464 & 0.641 \\
             $^{107}$Ru & 6.5 & 0.3 & 2.5 & 12.57 & 0.242 & -1.648\\
            $^{109}$Ru& 2.8 & 1.6 & 1.9 & 12.6 & 0.241 & -1.084 \\
             $^{111}$Ru & 8  & 0.1 & 2.1  & 13.01 & 0.237 & -0.713 \\
             $^{113}$Ru& 2 & 2.8 & 1.3 &  13.1 & 0.265 & 0.063 \\
           
\end{tabular}
\end{ruledtabular}
\end{table*}

\begin{table*}
\caption{\label{tab:rigid-params} Parameters of the rigid core calculations, where $\gamma$ is  the triaxiality parameter, $\kappa$ is the coupling strength of the quadrupole interaction, $E_{2}$ is the energy of the first $ 2^{+}$ state of the cores, and $\varepsilon_{j}-\lambda$ is the Fermi level relative to the neutron $h_{11/2}$ and the proton $g_{9/2}$  levels.}
\begin{ruledtabular}
\begin{tabular}{cccccc}
 Nucleus & $\gamma $ & c & $\kappa$& $E_{2}$ & $\lambda-\varepsilon_{j}$  \\ \hline
 $^{101}$Pd& 30 & 0.07 & 8.6 & 0.556 & -3.487 \\
  $^{103}$Pd & 30 & 0.07 & 9.11 & 0.556 & -2.96\\
  $^{105}$Pd& 30 & 0.07 & 9.92 & 0.512 & -2.47 \\
  $^{107}$Pd& 30& 0.07 & 10.46 & 0.434 & -2.028\\
  $^{109}$Pd & 30& 0.09 & 11 & 0.44 & -1.58 \\
   $^{111}$Pd& 30&  0.07  & 9.36 & 0.433 & -1.0482 \\
    $^{113}$Pd & 30& 0.07 & 7 & 0.415 & -0.62 \\
     $^{115}$Pd& 30& 0.07 & 8.7 & 0.475 & -0.184 \\
      $^{101}$Rh & 10&  0.09 & 8.6 & 0.5 & 0.61 \\
       $^{103}$Rh& 15& 0.09 & 9.11 & 0.556 & 0.63 \\
        $^{105}$Rh & 10& 0.07  & 9.92 & 0.44 & 0.716 \\
         $^{107}$Rh& 10& 0.07 & 10.46 & 0.434 & 0.76\\
          $^{109}$Rh & 14&  0.09 & 11 & 0.34 & 0.847 \\
           $^{111}$Rh& 14& 0.07  & 9.36 & 0.349 & 0.697 \\
            $^{113}$Rh & 11&  0.09  & 7 & 0.33 & 0.492 \\
             $^{115}$Rh& 11 & 0.08 & 8.7 & 0.34 & 0.641 \\
             $^{107}$Ru & 30 & 0.09  & 12.57 & 0.4 & -1.648 \\
              $^{109}$Ru& 44 &  0.09 & 12.6 & 0.4 & -1.084 \\
               $^{111}$Ru & 44& 0.09 & 13.01 & 0.4 & -0.713 \\
                $^{113}$Ru& 17&  0.09  & 13.1 & 0.3 & 0.063 \\
\end{tabular}
\end{ruledtabular}
\end{table*}

\subsection{Rigid core }
\label{sec:background:rigid}

For the rigid triaxial core, only the triaxial rotor Hamiltonian (\ref{Hrot}) is
taken into account. The triaxiality parameter $\gamma$ is fixed and adjusted as
specified below.  The MoI are taken to have irrotational flow dependence
(\ref{MoI}) on $\gamma$ and a scale which increases with core angular momentum
$R$ as
\begin{equation}\label{JR}
\mathcal{J}_{k}'(R)=\Theta(1+cR)\sin (\gamma -\frac{2}{3}\pi k)^{2}.
\end{equation}
The parametrization $\Theta(1+cR)$ for the scale factor is motivated by the
study of tidal waves in Ref.~\cite{FGS11} (see also
Ref.~\cite{background}).  There are two effects taken into account this way. The
Coriolis force acts on both spins of a pair of nucleons with angular momenta in
opposite directions, trying to align nucleons to the rotational axis as the
total angular momentum $R$ increases, which causes a reduction of the pair
correlations and an increase of the MoI~\cite{peter}. A second effect is a
stretching of the nucleus in the $\beta$ direction with increasing $R$ for a
potential with finite rigidness. The MoI of the Bohr Hamiltonian kinetic
energy~(\ref{MoI}) is proportional to $\beta^{2}$. Thus, the simple relationship
(\ref{JR}) between $\mathcal{J}'_{i}$ and $R$ is expected~\cite{background}.

Thus, the triaxial rotor core is not a \enquote{rigid} rotor in the literal
sense.  The excitation energies of the core decrease with $(1+cR)$, which will
influence the coupling between core and quasiparticle.  To keep terminology
simple, we will refer, somewhat loosly, to this triaxial rotor as the ``rigid''
core and to the collective model with the triaxial Davidson potential as the
``soft'' core.  The parameters of the rigid core are listed in
Table~\ref{tab:rigid-params}, where $\Theta$ is fixed by the energy $E(2^+_1)$
of the core.  The excitation energies $E_{c}(R) $ and $ \langle Rk \Vert Q \Vert
R'k'\rangle$ of the core are calculated by numerical diagonalization of the
triaxial rotor Hamiltonian (\ref{Hrot}).

\begin{figure}[t]
\begin{center}
\includegraphics[width=\hsize]{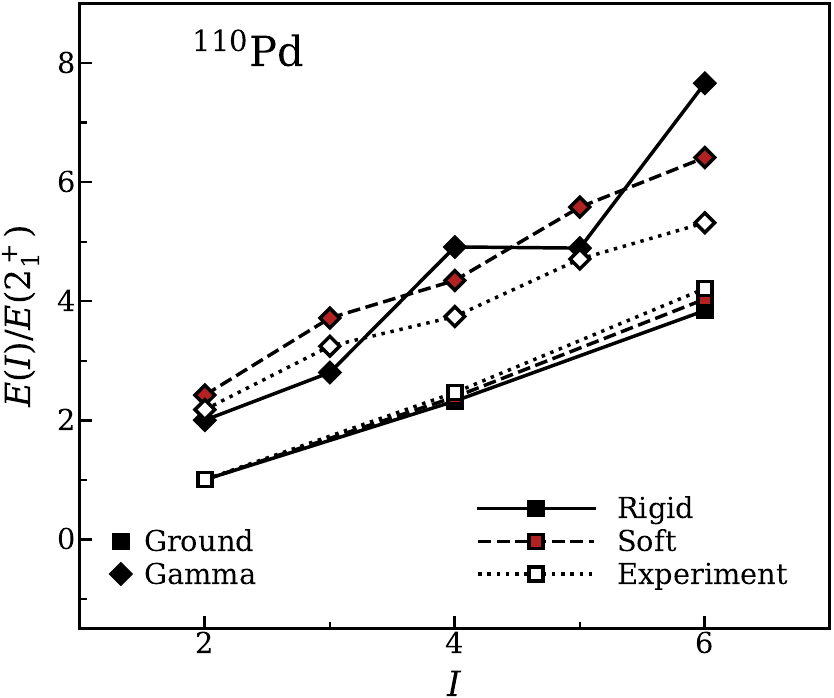}
\end{center}
\caption{Ground and $\gamma$ band of the soft collective model and rigid rotor
  core in comparison with experiment $^{110}$Pd.  The soft core represents a
  typical fit of the collective model with triaxial Davidson potential to the
  experiment (cf.~Table~\ref{tab:soft-params} for parameters), which is used in
  the QTD calculations.  The rigid rotor core is used in the QTR calculations
  (cf.~Table~\ref{tab:rigid-params} for parameters).
\label{fig:staggering}
}
\end{figure}

\subsection{Staggering of the $\gamma$ band}
\label{sec:background:staggering}

A sensitive indicator of the rigidity of the collective core is the energy
staggering of the $\gamma$ band, which is defined as~\cite{stagger}
\begin{equation}\label{eq:S}
  S(R)=\frac{E(R)-2E(R-1)+E(R-2)}{E(2_{1}^{+})},
\end{equation}
where the $E(R)$ are the energies for spin $R$ in the $\gamma$ band, and
$E(2_{1}^{+})$ is the energy of the first $2^{+}$ state in the ground band.  For
a $\gamma$-soft core, the energies of the even-$R$ levels lie below the average
of the adjacent odd-$R$ levels. For the $\gamma$-rigid core, the energies of the
odd-$R$ levels lie below the average of the adjacent even-$R$
levels~\cite{staggeringOddandEven}.

Fig.~\ref{fig:staggering} shows the ground band and $\gamma$ band of $^{110}$Pd as an example for a core. 
The ground band of the collective model core and the rigid core are close to experiment, while the $\gamma$ band of the rigid rotor 
has opposite staggering compared to both experiment and the collective model core.

\subsection{Core-Quasiparticle Coupling Model}
\label{sec:background:cqpc}
 
In the present work we use the core-quasiparticle coupling (CQPC)
model~\cite{cqpc}. CQPC generalizes the familiar concept of BCS quasiparticles
in a statically deformed potential by introducing an extended set which consists
of a multitude of quasiparticle sets, each of which belongs to a potential with
a different shape, where appropriate deformation parameters cover the range of
the collective motion of the core. The pair field is assumed to be independent
of the shape of the core.  The quasiparticle are coupled by the kinetic energy
operator of the collective core motion.

The CQPC model is formulated as an eigenvalue problem in the vector space where
a particle is coupled to a collective states of the $A-1$ neighbor as
$c^{+}_{j,m}|RM_{R}kA-1\rangle$ and a hole is coupled to the collective states
of the $A+1$ neighbor as $c^{-}_{j,m}|RM_{R}kA+1\rangle$. Here, $R$ is the
angular momentum of the core, $M_{R}$ is its projection and \textit{k}
summarizes all the remaining quantum numbers of the collective states of the
core.

The Hamiltonian for the odd-mass nucleus is given by
\begin{equation} 
  H=H_{af}+H_{c},
\end{equation}
where $H$ is a matrix in the above defined vector space. 
The matrix $H_{af}$ (adiabatic field) couples the particle and the hole to the instantaneous deformed potential and pair field,  and the matrix $H_{c}$ describes the collective 
motion of the  even-even core. The eigenvalues $E_{I}$ and the eigenvectors $\overrightarrow{uv_{I}}$ are found by diagonalizing  the matrix $H$:
\begin{equation}
  H\overrightarrow{uv_{I}}=E_{I}\overrightarrow{uv_{I}}.
\end{equation}
The eigenvalues of this Hamiltonian are the excitation energies of the odd-$A$ nucleus. 
The eigenvectors are
\begin{equation}
 \overrightarrow{uv_{I}} =\begin{bmatrix}
                            \overrightarrow{u_{I}} \\
                           \overrightarrow{v_{I}}
                            \end{bmatrix}=
 			\begin{bmatrix}
                            \langle RM_{R}kA-1|c_{jm}|IMA \rangle \\
                            \langle RM_{R}kA+1|c^{+}_{j\bar{m}}|IMA \rangle
                            \end{bmatrix},
                         \end{equation}                           
 where $ \overrightarrow{u_{I}}$ represents the probability amplitude for the considered states $|IMA\rangle$ of the odd-$A$ nucleus composed of a particle in the (A-1) nucleus, 
 and $ \overrightarrow{v_{I}}$ represents the probability amplitude for a hole in the (A+1) nucleus. 
 Here, \textit{jm} are the quantum numbers of the particle and hole within a major shell, $RM_{R}k$ are the 
 quantum numbers of the collective states of the even-even core, and $IM$ are the angular momentum and its projection of the odd mass nucleus.
 Note, the square of the amplitudes are the spectroscopic factors for particle transfer reactions.  
 
The Hamiltonian of the odd quasiparticle reads 
\begin{equation}
H_{af}=\begin{bmatrix}
\varepsilon -\lambda +\tau  & \Delta \\ 
 \Delta &   -(\varepsilon -\lambda +\tau)
\end{bmatrix},
\end{equation}
and the Hamiltonian of the even-even core
\begin{equation}
H_{c}=\begin{bmatrix}
       E_{c} & 0 \\ 
         0   &  E_{c}
       \end{bmatrix}.
\end{equation}
In these equations, $\varepsilon_j$ is the energy of the spherical $j$-shell, $\lambda$ is the chemical potential, $\Delta$ is the pairing field, 
and $E_{c}$ are the energies of the core collective quadrupole excitations.

 The interaction  $\tau$ 
between the particle and the core is described by the quadrupole field, which can be expressed as:
 \begin{eqnarray}
(\tau )_{jRk,j'R'k'}&=&-\frac{\kappa}{2}(-1)^{j'+R'+I}\begin{Bmatrix}
j &j'  &2 \\ 
 R& R' &I
\end{Bmatrix}\nonumber \\&& \times \langle R k\Vert Q\Vert R' k' \rangle  q_{jj'} ,
\end{eqnarray} where $ \langle R k\Vert Q\Vert R'k'\rangle $ and $ q_{jj'}$ are the quadrupole reduced matrix elements of the core and particle, and $\kappa$ is the coupling strength of the quadrupole field.
The quantum numbers of the even-even core are $R$, $k$, $R'$, $k'$, and $j$, $j'$ are quantum numbers of the particle. 
The $M_{R}$ and $m$ dependence is taken care of by angular momentum algebra, which leads to the reduced matrix elements~\cite{cqpc}.
 
In the general application of the QPCM the spherical single particle energies
$\varepsilon_{j}$ are taken from a spherical potential, as the Nilsson or
Woods-Saxon potential or some more sophisticated energy density functional (EDF)
approach. In the present paper only the coupling of the $h_{11/2}$ quasineutrons
and the $g_{9/2}$ quasiprotons is studied.  For them one can set
$\varepsilon_{j}=0$, without loss of generality.  The pairing gap $\Delta$ is
set to 1.226 MeV.

The coupling strength of the quadrupole interaction $\kappa$ is evaluated using
the self-consistency condition for the QQ-model
\begin{equation}
  \kappa/5=\hbar w_{0}\bar{\beta} \langle 2 \Vert Q \Vert 0 \rangle, 
\end{equation}
where $\hbar w_{0}=41A^{-1/3}$MeV, and the scale $Q_0$ of the core matrix elements is chosen such that $\langle 2 \Vert Q \Vert 0 \rangle=1$. 
The core deformation $\bar{\beta}$ is taken from the compilation \cite{raman}, 
which relates  $\bar{\beta}$ to the experimental $B(E2;0\rightarrow2)$ by 
\begin{equation}
  \bar\beta=\left[\frac{4\pi Ze^2R_{0}^{2}}{3B(E2;0\rightarrow2)}\right]^{1/2}.
\end{equation} 
The resulting coupling constants $\kappa$ are listed in
Tables~\ref{tab:soft-params} and~\ref{tab:rigid-params}.

The particle number \textit{n} for the state $\vert IMA\rangle$ is given by
\begin{equation}\label{eq:n}
  n=\sum_{jmkRM_{R}}| \langle IMA|c^{+}_{jm}|RM_{R}kA-1 \rangle |^{2}.
\end{equation}
As  usual in BCS, the chemical potential $\lambda$ is determined by the condition $n=A$. In practice, it is fixed for the ground state of the odd nucleus
and kept constant for the other states.  
In this paper we only study the coupling to the  $h_{11/2}$ neutron and the $g_{9/2}$ protons, which are approximated as pure $j$-shells. In the calculations 
$\lambda-\varepsilon_j$ is needed, which is also listed in the tables. 
All single 
particle energies $\varepsilon_{j}$ within the major shell are needed to obtain the  particle number (\ref{eq:n}), which are taken from the Nilsson potential.

\subsection{Wobbling motion of triaxial nuclei}
\label{sec:background:wobbling}

\begin{figure*}[t]
 \begin{center}
    \includegraphics[width=\hsize]{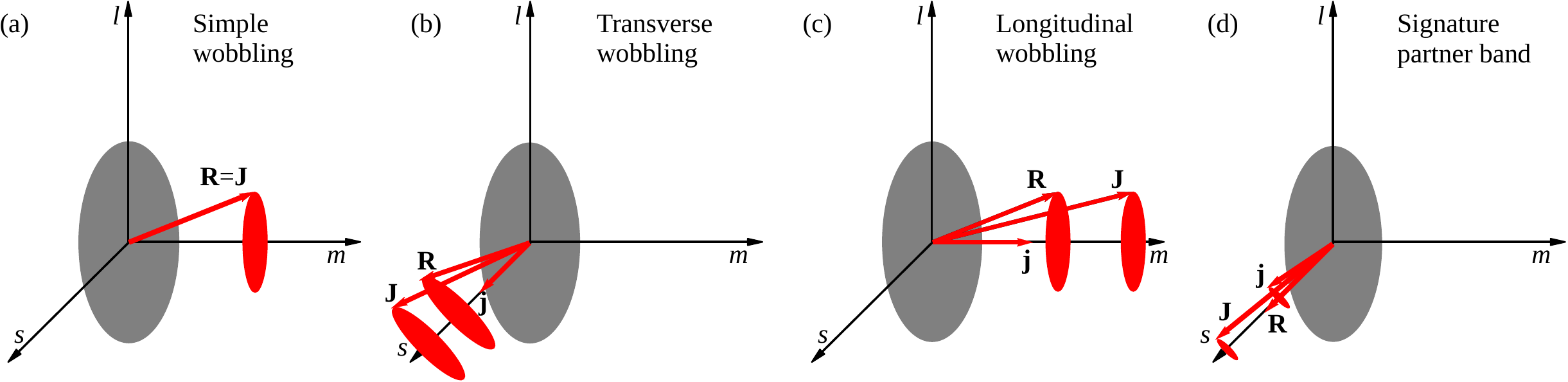}
 \end{center}
 \caption{Different types of wobbling motion. The four panels correspond to
      (a)~simple wobbbling, (b)~transverse
      wobbling, (c)~longitudinal wobbling, and (d)~the signature partner band. Here $l$, $m$ and $s$
      indicate the long, medium and short axes, respectively, with corresponding
      MoI ordered as
      $\mathcal{J}_{m}>\mathcal{J}_{s}>\mathcal{J}_{l}$. The arrows represent the
      angular momentum $\vec{R}$ of the core, the angular momentum $\vec{j}$ of the quasiparticle, and the total angular momentum $\vec{J}$.
      }
    \label{fig:wobbling}
\end{figure*}

The coupling of an odd quasiparticle to a rigid triaxial rotor has been
classified in Ref.~\cite{transverse}.  This classification is summarized in
Fig.~\ref{fig:wobbling}, where the axes represent the medium ($m$), short ($s$), 
and long ($l$) body-fixed axes of the nucleus.  The MoI are assumed to be
ordered as  $\mathcal{J}_{m}>\mathcal{J}_{s}>\mathcal{J}_{l}$ for these axes,
respectively.

The wobbling motion of the triaxial rotor itself, without an odd quasiparticle,
has been discussed by Bohr and Mottelson \cite{BohrMottlelsonII}.  It is
illustrated in Fig.~\ref{fig:wobbling}(a), where, following
Ref.~\cite{transverse}, such wobbling is termed ``simple wobbling''. In
classical mechanics, uniform rotation around the axis with the largest MoI
(here, the $m$-axis) has minimal energy for given angular momentum. At larger
energy, rotation about the other two axes appears, which generates a wobbling of
the triaxial body with respect to the space-fixed angular momentum vector.  In
the body-fixed frame, the wobbling motion represents a precession of the angular
momentum vector around the axis with the largest MoI (again, here, the
$m$-axis).
  
In the small-amplitude limit, such motion has a quantized energy spectrum given by:
\begin{equation}\label{swobe}
 \textit{E}(n_{w},I)=\frac{\hbar^2}{2\mathcal{J}_{m}}I(I+1)+(n_{w}+1/2)\hbar \omega_{w},
\end{equation}
where $n_{w}$ describes the wobbling phonon number and $I$ is the total
angular momentum~\cite{transverse}. The wobbling frequency $\omega_{w}$ is
determined by
\begin{equation}\label{swobom}
\hbar \omega_{w}=\frac{I\hbar^2}{\mathcal{J}_{m}}\sqrt{\left(\frac{\mathcal{J}_{m}}{\mathcal{J}_{s}}-1\right)\left(\frac{\mathcal{J}_{m}}{\mathcal{J}_{l}}-1\right)}.
\end{equation}
 It is to be compared with the experimental
wobbling energy $\textit{E}_{w}$ defined as~\cite{transverse}:
\begin{eqnarray}\label{ewob}
\textit{E}_{w}(I) &=& \textit{E}(n_{w}=1,I)\\&&-[\textit{E}(n_{w}=0,I-1)+\textit{E}(n_{w}=0,I+1)]/2,\nonumber
\end{eqnarray}
For an even-even triaxial nucleus or \enquote{simple wobbler}, the wobbling energy increases with $I$.

The wobbling motion is modified for odd-mass nuclei, as discussed in
Ref.~\cite{transverse}.  Fig.~\ref{fig:wobbling}(b) illustrates the case where there is a
high-$j$ quasiparticle coupled to the core.
The
quasiparticle's angular momentum $j$ aligns with the $s$-axis (with intermediate MoI),
which is transverse to the $m$-axis (with the maximal MoI).  The wobbling motion
consists in the precession of the angular momentum vector about the transverse
$s$-axis and is thus termed ``transverse wobbling''~\cite{transverse}. In
the harmonic approximation, the energy of the wobbling bands is given by
\begin{equation}\label{twobe}
 \textit{E}(n_{w},I)=\frac{\hbar^2}{2\mathcal{J}_{s}}(I+1/2-j)^2+(n_{w}+1/2)\hbar \omega_{w},
\end{equation} 
with the wobbling frequency 
\begin{equation}\label{twobom}
\hbar \omega_{w}=\frac{I\hbar^2}{\mathcal{J}_{s}}\sqrt{\left(\frac{\mathcal{J}_{s}}{\mathcal{J}_{m}}-1+\frac{j}{I}\right)\left(\frac{\mathcal{J}_{s}}{\mathcal{J}_{l}}-1+\frac{j}{I}\right)}.
\end{equation}
As $\mathcal{J}_{s}/\mathcal{J}_{m}<1$, the first factor $
({\mathcal{J}_{s}}/{\mathcal{J}_{m}}-1+{j}/{I})$ inside the radical decreases
with $I$, until it crosses zero,
while, as $\mathcal{J}_{s}/\mathcal{J}_{l}>1$, the second factor $
({\mathcal{J}_{s}}/{\mathcal{J}_{l}}-1+{j}/{I})$ monotonically decreases.
Consequently, considering the effects of all factors together, the wobbling frequency $\hbar \omega_{w}$ first briefly increases with
$I$, then decreases with $I$, until it vanishes at the critical angular momentum
\begin{equation}
 I_c=\frac{j \mathcal{J}_{m}}{\mathcal{J}_{m}-\mathcal{J}_{s}},
\end{equation}
at which point the transverse wobbling
mode becomes unstable.  This is to be contrasted with the simple wobbler above, for which
the wobbling frequency~(\ref{swobom}) simply increases linearly with $I$.
 
The case when the angular momentum $j$ of the quasiparticle is aligned with the
$m$-axis, which appears for a quasiparticle in a half-filled $j$-shell, is
illustrated in Fig.~\ref{fig:wobbling}(c).  The wobbling motion then consists in
the precession of the angular momentum vector about the $m$-axis (with the
maximal MoI) and is thus termed ``longitudinal wobbling''~\cite{transverse}.  In
the harmonic approximation, the energy of the wobbling bands is given by
\begin{equation}\label{lwobe}
 \textit{E}(n_{w},I)=\frac{\hbar^2}{2\mathcal{J}_{m}}(I+1/2-j)^2+(n_{w}+1/2)\hbar \omega_{w},
\end{equation} 
with the wobbling frequency 
\begin{equation}\label{lwobom}
\hbar \omega_{w}=\frac{I\hbar^2}{\mathcal{J}_{m}}\sqrt{\left(\frac{\mathcal{J}_{m}}{\mathcal{J}_{s}}-1+\frac{j}{I}\right)\left(\frac{\mathcal{J}_{m}}{\mathcal{J}_{l}}-1+\frac{j}{I}\right)}.
\end{equation}
As $\mathcal{J}_{m}/\mathcal{J}_{s}>1$ and $\mathcal{J}_{m}/\mathcal{J}_{l}>1$,
both factors within the radical decrease with $I$.  This has the consequence
that, although the frequency $\hbar w_{w}$ of longitudinal wobbling increases
with $I$, it does so less rapidly than for the simple
wobbler~(\ref{swobom}).
  
In the following we call the lowest-energy band with $n_{w}=0$ the ``yrast
band'', and we call the one-phonon ($n_{w}=1$) excited band on top of the yrast
band the ``wobbling band''. Because of the $D_{2}$-symmetry of the triaxial
potential the bands have good signature $\alpha=\pm1/2$.  They form $\Delta I=2$
sequences with $I=1/2,5/2,9/2,\ldots$ for $\alpha=+1/2$ or
$I=3/2,7/2,11/2,\ldots$ for $\alpha=-1/2$. For the considered case of $h_{11/2}$
quasiparticles, the yrast band has $\alpha =-1/2$, and the wobbling band has
$\alpha =+1/2$.

There is another $\alpha =+1/2$ band, which we call the ``signature partner''
band.  The signature partner for the case of transverse coupling of the
quasiparticle is illustrated in Fig.~\ref{fig:wobbling}(d).  For the signature
partner band, the angular momentum of the quasiparticle is no longer maximally
aligned ($m=j$) with the $s$-axis, but it precesses around the $s$-axis with
$m=j-1$, where the collective angular momentum is aligned with the $s$-axis.
Signature partner bands are well known from the classification of rotational
spectra in terms of quasiparticles in a rotating potential \cite{peter}.

As an example, Fig.~\ref{fig:109pd}(b) below (in Sec.~\ref{sec:discussion}),
shows the three lowest bands in $^{109}$Pd obtained by the QTR calculation that
couples a $h_{11/2}$ quasineutron to a triaxial rotor with
$\gamma=30^\circ$. The lowest band, which we refer to as yrast
in the following, has $\alpha=-1/2$. The two excited bands have $\alpha=1/2$.

The probabilities for the $\Delta I=1$ transitions to the yrast band clearly
distinguish the wobbling band from the signature partner band.  As seen in
Fig.~\ref{fig:109pd}(d), the $B(E2)$ values of the transitions from the band
that we classify as wobbling are much larger than those from the signature
partner band.  As illustrated in Fig.~\ref{fig:wobbling}(b), the precession of
the total angular momentum vector $\vec J$ about the $s$-axis of the body fixed
frame means that the triaxial charge distribution wobbles in the laboratory
system, which generates collectively enhanced $E2$ radiation.  By analyzing the wave
functions of the odd-mass nuclei, we found that for the same total $J$, the core
angular momentum $\vec{R}_{\perp}$ of the wobbling band is large. This is
because transverse wobbling approaches instability with respect to a permanent
tilt of the rotational axis relative to the $s$-axis. The cases with
longitudinal wobbling character have been found to have a large wobbling amplitude as
well.

To discuss the signature partner band it is noted that the total angular
momentum is expressed as $\vec{J}=\vec{j}+\vec{R}$, and its transverse component
relative to the total angular momentum $\vec{J}_{\perp}=0$ by definition, which
gives $\vec{R}_{\perp}=-\vec{j}_{\perp}$. As illustrated in
Fig.~\ref{fig:wobbling}(d), the quasiparticle is no longer aligned with the
$s$-axis, but precesses around it with $j_{\perp}=\sqrt{2j}\approx 2$.  The
associated precession of $\vec J$ corresponds to much weaker oscillations of the
charge quadrupole moment than the large-amplitude precession of $\vec J$ of the
wobbling band. The oscillations of the quadrupole moment generate the $E2$
radiation, where $B(E2)$ is proportional to the squared amplitude of these
oscillations~\cite{transverse}, which explains the large difference between the
$B(E2)$ values.

Fig.~\ref{fig:109pd}(e) shows that the $B(M1)$ values of the transitions from
band that we classify as wobbling are much larger than those from the
signature band.  The oscillations of the magnetic moment generate the $M1$
radiation, where $B(M1)$ is proportional to the squared amplitude of these
oscillations~\cite{transverse}.  The magnetic moment is given by
\begin{equation}
  \vec{\mu}_{\mathrm{total}}=\vec{\mu}_{\mathrm{particle}}+\vec{\mu}_{\mathrm{core}},
\end{equation}
with
\begin{equation}
  \vec{\mu}_{\mathrm{particle}}= g_j \vec{j} \quad \vec{\mu}_{\mathrm{core}}= g_R \vec{R},
\end{equation}
where $g_j$ and $g_R$ are the single-particle and core $g$ factors, respectively.
As the vectors $\vec{j}$ and $\vec{R}$ precess
about the constant total angular momentum $\vec{J}$, only $\vec{R}_{\perp}$ and
$\vec{j}_{\perp}$ are time dependent. The transverse component of the total
magnetic moment can be therefore written in either of two ways:
\begin{align}\label{muj}
\vec{\mu}_{\perp}&=\vec{j}_{\perp}(g_j-g_R)
\\
\label{muR}
\vec{\mu}_{\perp}&=\vec{R}_{\perp}(g_R-g_j).
\end{align}
The geometry of the transverse
wobbling band is illustrated in Fig.~\ref{fig:wobbling}(b).  Since the
quasiparticle is aligned with $s$-axis when the $M1$ transitions from the
wobbling band to the yrast band are considered, we use the
expression~(\ref{muR}) to describe the transverse magnetic moment. As discussed
for the $E2$ transitions, $\vec{R}_{\perp}$ of the wobbling band is large,
because transverse wobbling approaches instability with respect to a permanent
tilt of the rotational axis relative to the $s$-axis.
  
For the magnetic transitions from the signature partner to the yrast band we use
expression~(\ref{muj}).  As illustrated in Fig.~\ref{fig:wobbling}(d) and discussed for
the $E2$ transition, the quasiparticle precesses around the m-axis, with
$j_{\perp}=\sqrt{2j}\approx 2$, which is much smaller than $R_{\perp}$ for the
wobbling band. Accordingly, the $B(M1)$ of the transition from the wobbling band
to the yrast band is much bigger than that of the transition from the signature
partner band to the yrast band.

\section{Discussion}
\label{sec:discussion}

\begin{figure*}[t]
\begin{center}
\includegraphics[width=\hsize]{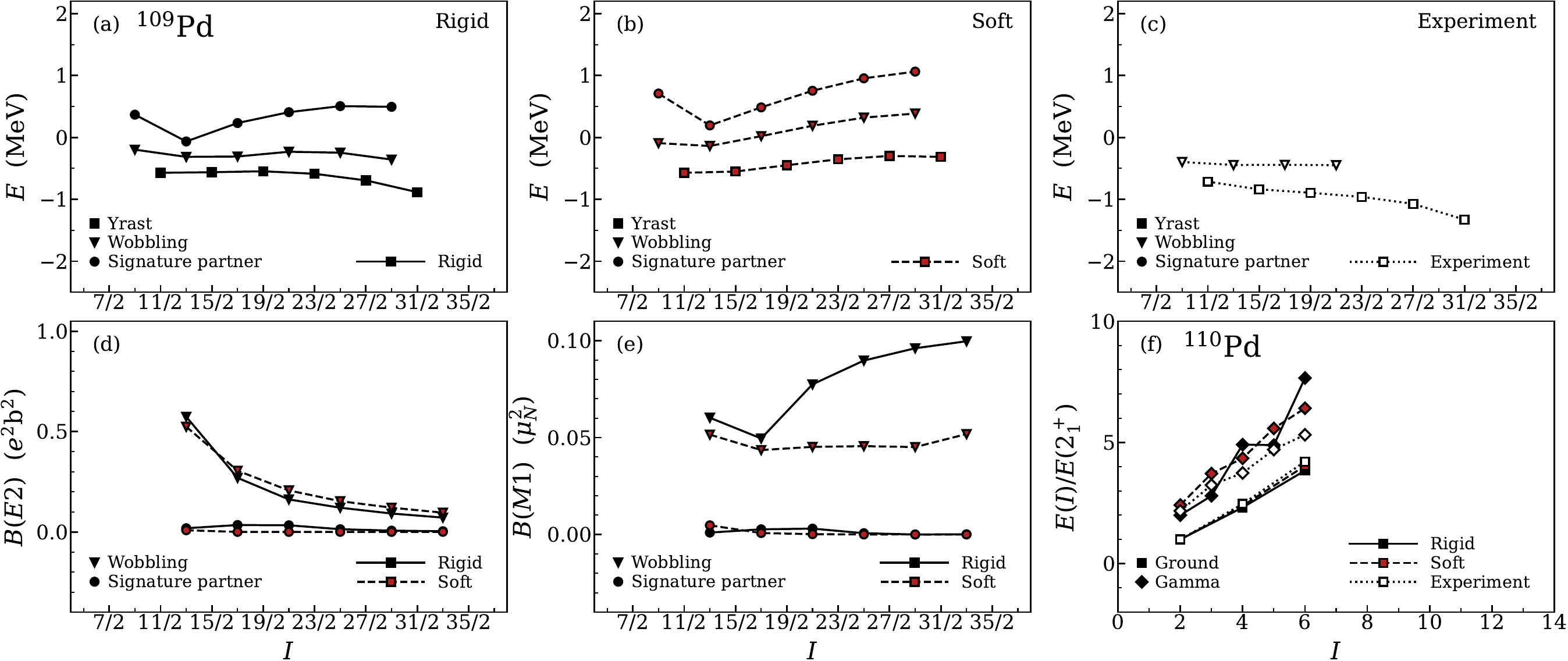}
\end{center}
\caption{Comparison between QTR  (rigid core)  and  QTD (soft core) predictions, along with
  experiment, for $^{109}$Pd: (a-c) Energies, minus a rotor contribution, within
  the yrast band (squares), wobbling band (triangles), and signature partner
  band (circles).  (d-e) Transition $B(E2)$ and $B(M1)$ strengths from the
  wobbling band (triangles) and signature partner band (circles)  to the yrast
  band.  (f) Normalized energies within the ground band (squares) and $\gamma$
  band (diamonds) for the $^{110}$Pd core.  Observables are shown
  vs.\ $I$, for rigid calculations (solid symbols, solid lines), soft
  calculations (red shaded symbols, dashed lines), and experiment (open symbols,
  dotted lines).
\label{fig:109pd}
  }
\end{figure*}

\begin{figure*}[t]
\begin{center}
\includegraphics[width=\hsize]{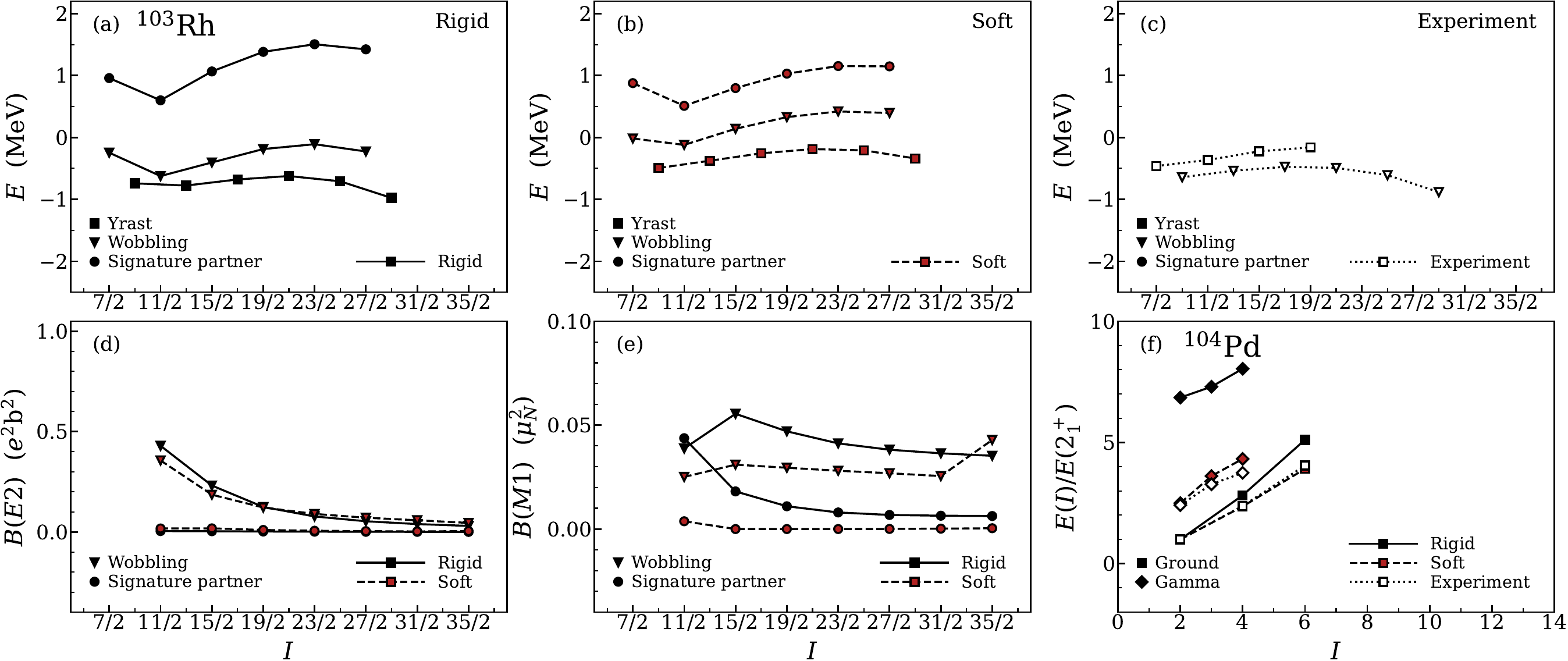}
\end{center}
\caption{Comparison between QTR (rigid core) predictions, QTD (soft core)
  predictions, and experiment for $^{103}$Rh, along with normalized energies
  for the $^{104}$Pd core. See Fig.~\ref{fig:109pd} caption for description of
  observables and notation.
\label{fig:103rh}
}
\end{figure*}

\begin{figure*}[t]
\begin{center}
\includegraphics[width=\hsize]{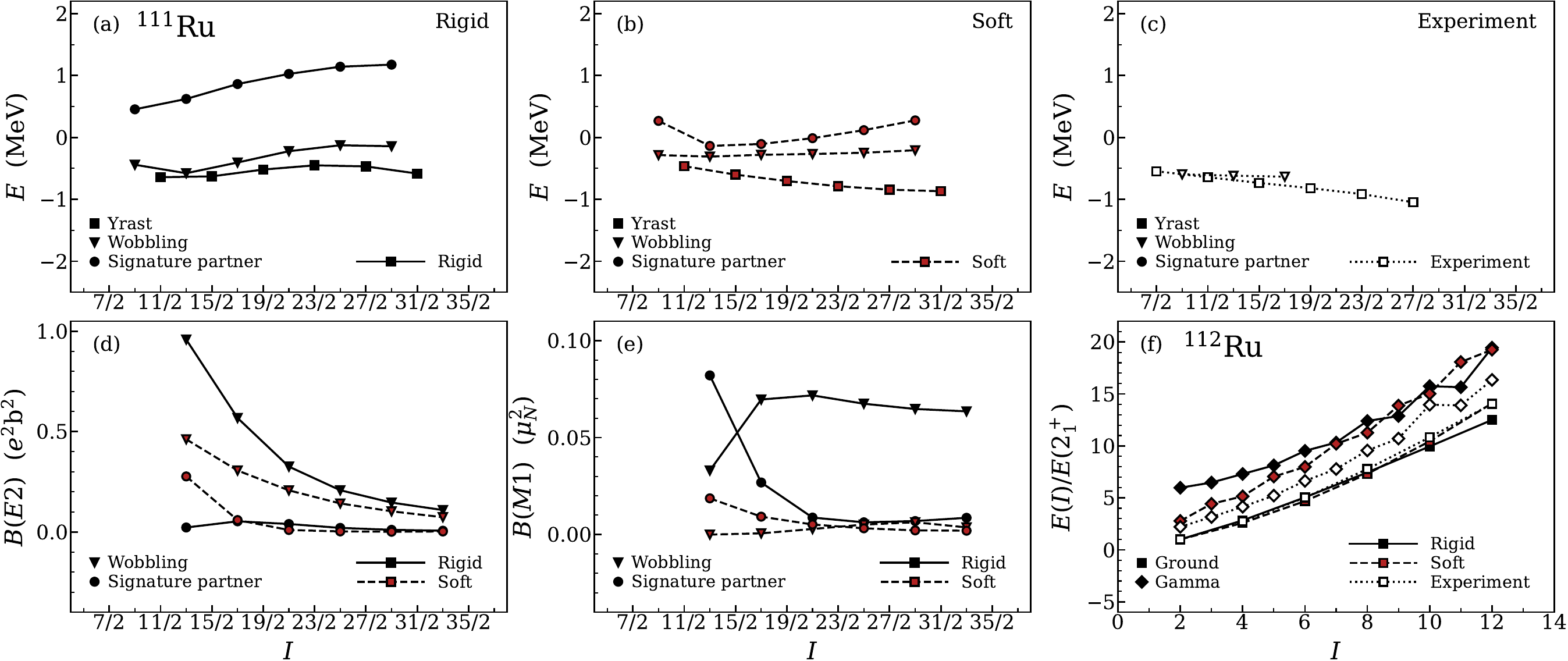}
\end{center}
\caption{Comparison between QTR (rigid core) predictions, QTD (soft core)
  predictions, and experiment for $^{111}$Ru, along with normalized energies
  for the $^{112}$Ru core. See Fig.~\ref{fig:109pd} caption for description of
  observables and notation.
\label{fig:111ru}
}
\end{figure*}

The purpose of this paper is to compare the coupling of a quasiparticle to the
collective model core with soft triaxial potential, through quasipartical
triaxial Davidson (QTD) calculations which take fluctuation of the triaxial
shape explicitly into account, and quasiparticle triaxial rotor (QTR)
calculations, which do not do so.  
The consequences of triaxiality for an odd-mass nucleus will be studied by
comparing the two different approaches as follows. In the
QTR calculations the $\gamma$ parameter of the TR is adjusted such that the
calculated wobbling energy of odd-mass nucleus is obtained as close as possible
to the experimental wobbling energy. In the QTD calculations, the properties of
even-even nuclei are fixed first with experimental bands energies, then a
quasiparticle is coupled to the core to describe odd mass nuclei.
The model parameters used for calculating
properties of $^{109-115}$Pd, $^{109-115}$Rh and $^{107-113}$Ru are listed in
Table~\ref{tab:soft-params} for the soft core and in
Table~\ref{tab:rigid-params} for the hard core.

Fig.~\ref{fig:109pd}(a)-(c) show the experimental and calculated (QTR and QTD)
level energies for yrast, wobbling and signature partner bands in
$^{109}$Pd. For $N=63$ the occupation of the $h_{11/2}$ shell is about
3. Accordingly, Fig.~\ref{fig:109pd}(a) shows that the QTR starts with a
decreasing wobbling energy as a function of angular momentum. This transverse
wobbling regime quickly becomes unstable and changes into the longitudinal
wobbling regime with increasing wobbling energy (see discussion of the
stability of the transverse wobbling regime in Ref.~\cite{Chen22}). The QTD
calculation in Fig.~\ref{fig:109pd}(b) shows a steady increase of the wobbling
energy with $I$, which signals the longitudinal wobbling regime. The
fluctuations of $\gamma$ toward axial shape further weaken the already weak
transverse coupling of the quasineutron, which results in longitudinal wobbling
wobbling right from the lowest values of $I$. This is consistent with the steady
increase seen in experiment in Fig.~\ref{fig:109pd}(c) and the missing
staggering of the $\gamma$ band of $^{110}$Pd in Fig.~\ref{fig:109pd}(f).
  
Fig.~\ref{fig:109pd}(d) and (e) compare the electromagnetic ($E2$ and $M1$)
reduced transition probabilities.  The $B(E2)$ values are quite similar between
the approaches, whereas the $B(M1)$ values of QTR for the transitions from
wobbling to yrast band are higher than those of QTD. Fig.~\ref{fig:109pd}(f)
shows the energy ratios vs.\ $I$ of the ground band and $\gamma$ band of the core
$^{110}$Pd for the experiment, collective model and triaxial rotor.  The TD
core, which is fitted to experimental energies, reproduces the ``even-$I$-low
pattern'' of the quasi-$\gamma$ band (the energies of even angular momentum
states are lower than that average of their odd angular momentum neighbors),
which is observed in $^{110}$Pd. The TR core parameters are adjusted to the
experimental wobbling band energies of $^{109}$Pd. The inherent assumption of
stable triaxiality is reflected by the characteristic ``odd-$I$-low pattern''
(the energies of odd angular momentum states are lower that that average of
their even angular momentum neighbors) of the quasi-$\gamma$ band of the core
\cite{background}.

Fig.~\ref{fig:103rh} provides the same information for $^{103}$Rh, which has a
quasi-proton coupled to $^{104}$Pd. The proton number in the $g_{9/2}$ shell is
around 6, which means the shell is about half-occupied. Accordingly, the
wobbling energy increases with $I$, where there is no qualitative difference
between the soft and rigid cores. Longitudinal wobbling appears for a
quasiparticle in a half-filled shell, which is longitudinally coupled.

Fig.~\ref{fig:111ru} shows $^{111}$Ru which has a quasi-neutron coupled to
$^{112}$Ru. For $N=68$ the $h_{11/2}$ shell is half-occupied. Accordingly, the
quasineutron is longitudinally coupled to the triaxial core, which in the QTR
calculation gives an increase in wobbling energy with $I$ that signals
longitudinal wobbling.  The QTD calculation gives quite similar results, which
is expected because the TD core fitted to the experiment comes close to the TR
core [see Fig.~\ref{fig:111ru}(f)].

The examples in Figs.~\ref{fig:109pd}, \ref{fig:103rh}, \ref{fig:111ru}
demonstrate that including softness by means of the phenomenological core like
the TD always leads to the longitudinal wobbling pattern of wobbling energy
increasing with $I$. Only if the quasiparticle has particle-like character
($^{109}$Pd) the rigid TR core generates the transient transverse wobbling
pattern, the presence or absence of chich can be taken as a qualitative measure
of the core's softness.

The $B(M1)$ values for transitions from the wobbling to the yrast band are a
more sensitive way to differentiate between the QTR and QTD calculations.  The
difference can be explained as follows.  The coupling between the quasiparticle
and the core is not rigid. Therefore, the total magnetic moment does not
rigorously follow the motion of the core. The strength of the coupling between
the quasiparticle and core will influence the amplitude of the oscillations of
the magnetic moment. When the core includes $\beta$- and $\gamma$-vibrational
motion, these fluctuations will reduce the coupling strength between the
quasiparticle and the core. This is in contrast to the rigid TR core, which has
only rotational motion.  This explains why the $B(M1)$ of the transitions from
the wobbling to the yrast band are bigger in the QTR calculations than in the
QTD calculations.
 
\begin{figure*}[t]
\begin{center}
\includegraphics[width=0.70\hsize]{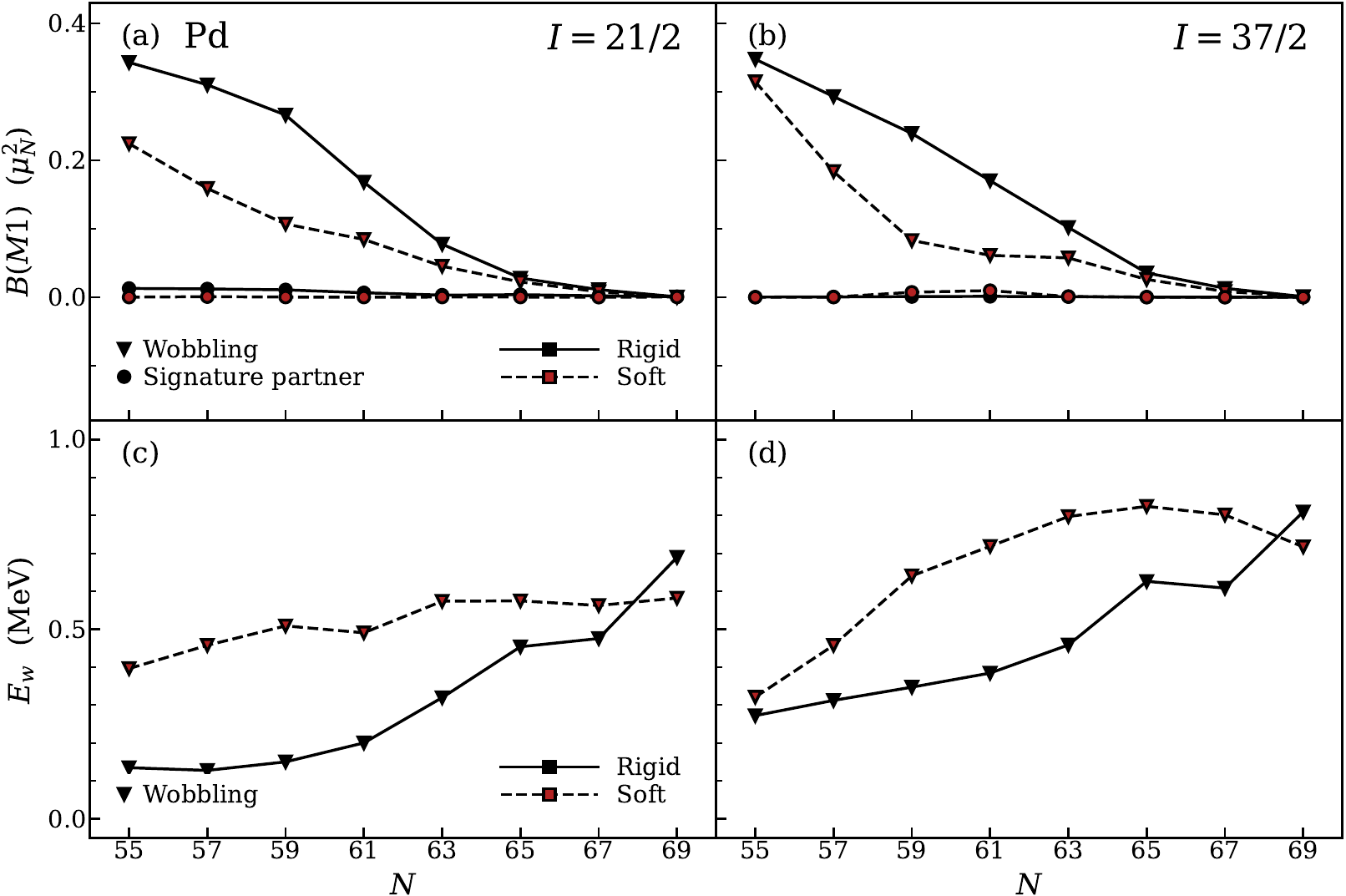}
\end{center}
\caption{%
  Comparison between QTR (rigid core) predictions and QTD (soft core)
  predictions for the Pd isotopic chain: (top)~$B(M1)$ strength and
  (bottom)~wobbling energy $E_w$, at fixed angular momenta
  $I=21/2$~(left) and $I=37/2$~(right).
  Observables are shown for rigid calculations (solid symbols, solid lines) and soft
  calculations (red shaded symbols, dashed lines), and for the wobbling band
  (triangles) and (in the case of the $M1$ strength) signature partner band (circles).
  The $M1$ transition is for
  angular momenta $I\rightarrow (I-1)$, from the wobbling or signature partner
  band to the yrast band.
      \label{fig:Pd}
      }%
\end{figure*}

\begin{figure*}[t]
\begin{center}
\includegraphics[width=0.70\hsize]{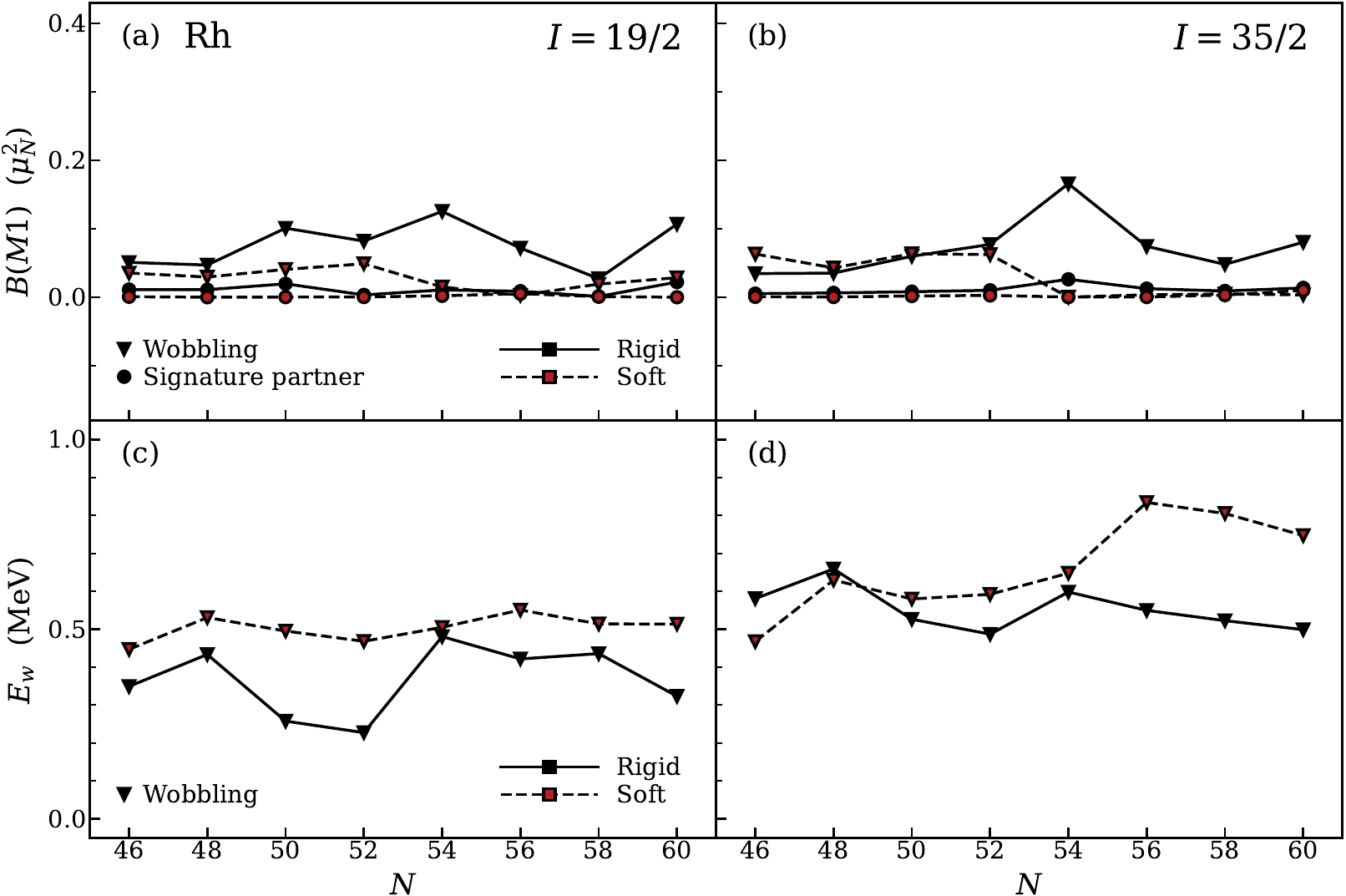}
\end{center}
\caption{%
Comparison between QTR (rigid core) predictions and QTD (soft core)
  predictions for the Rh isotopic chain: (top)~$B(M1)$ strength and
  (bottom)~wobbling energy $E_w$, at fixed angular momenta
  $I=19/2$~(left) and $I=35/2$~(right).
  See Fig.~\ref{fig:Pd} caption for description of
  observables and notation.
      \label{fig:Rh}%
      }
\end{figure*}

\begin{figure*}[t]
\begin{center}
\includegraphics[width=0.70\hsize]{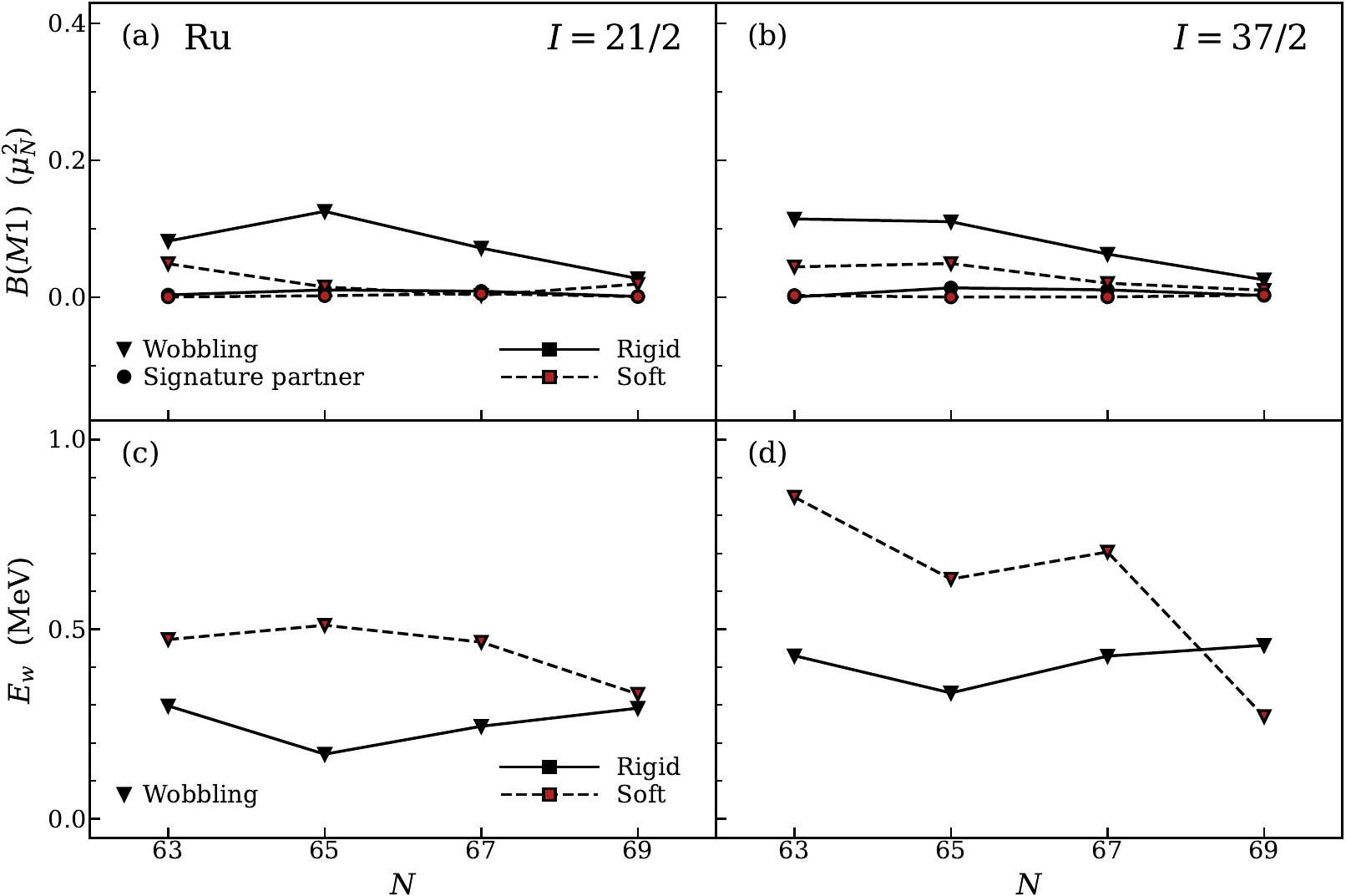}
\end{center}
\caption{%
Comparison between QTR (rigid core) predictions and QTD (soft core)
  predictions for the Ru isotopic chain: (top)~$B(M1)$ strength and
  (bottom)~wobbling energy $E_w$, at fixed angular momenta
  $I=21/2$~(left) and $I=37/2$~(right).
  See Fig.~\ref{fig:Pd} caption for description of
  observables and notation.
      \label{fig:Ru}
      }
\end{figure*}

Figs.~\ref{fig:Pd}, \ref{fig:Rh}, \ref{fig:Ru} compare QTR with QTD calculations
for the Pd, Rh, and Ru isoptopes, respectively. Only the results for two
selected values of $I$ are shown.  More detailed results can be found in
Ref.~\cite{thesisLi}.

Fig.~\ref{fig:Pd} shows the Pd chain.  For $I=21/2$ it can be seen that $B(M1)$
values for the transition from the wobbling to the yrast band decrease as the
neutron number increases, while the wobbling energies increase. This is
explained by the number of particles located in the $h_{11/2}$ shell, which
increases from 0.62 to 5.6 (half-occupied $h_{11/2}$ shell). So the
quasiparticle character changes half-way from particle-like towards hole-like as
the particle number in the $h_{11/2}$ shell increases. For the rigid TR core
this means the following. When the quasi-particle has a particle character, its
angular momentum aligns with the $s$-axis of the TR core, since in this way the
density distribution of the particle has the maximal overlap with core the
density distribution. When the quasi-particle has hole character, its angular
momentum aligns with the $l$-axis of the TR core, because for this orientation
the density distribution of the hole has minimal overlap with the core density
distribution.  When the quasiparticle' s character is in-between particle and
hole, its angular momentum tends to be aligned with the $m$-axis of the TR
core. When the quasiparticle is a particle or hole, it will oscillate rigidly
with the axis, giving a strong $B(M1)$. But when the quasiparticle is in-between
particle and hole, it will be partially decoupled from the core, resulting in a
small $B(M1)$ and a big wobbling energy. In contrast,the $E2$ radiation mostly
comes from the collective core, which is why the $B(E2)$ values do not change
dramatically with neutron number.  For the soft TD core the fluctuations away
from the triaxial shape attenuate the transverse coupling of the quasiparticle,
which is seen as larger wobbling energies and smaller $B(M1)$ values.  The
soft core and rigid core results approach each other near $N=66$ because the TD
fits approach the rigid rotor.
 
For $I=37/2$, the TR wobbling energies are about the same or larger than those
for $I=27/2$.  The detailed calculations \cite{thesisLi} show that the wobbling
energies first go down and then go up, like in Fig.~\ref{fig:109pd}, which
corresponds to a short transverse wobbling phase that changes into longitudinal
wobbling.  The TD wobbling energies for $I=37/2$ are larger than for $I=27/2$,
which signals the presence of the longitudinal wobbling regime as in
Fig.~\ref{fig:109pd}. The $B(M1)$ values for TD are smaller than those for
TR. The $I$ dependence is similar to that for $I=27/2$.
 
Fig.~\ref{fig:Rh} presents $B(M1)$ values and wobbling energies of the isotopes
$^{101-115}$Rh.  The proton number in the $g_{9/2}$ shell is constant around 6.
As a consequence, the character of the quasiparticle does not change. The
changes of the wobbling energies and $B(M1)$ values are caused by the
$N$-dependence of the core.  The dip of the QTR wobbling energy around $N=50$
may be an artifact, because modeling these nuclei as TR is not very realistic.
The more realistic TD cores give a smooth $N$-dependence in the region. For
$N\ge 58$ the QTR model seems acceptable. The comparison with the QTD shows the
same pattern as for the odd-neutron nuclei. The QTD wobbling energies are larger
than the QTR ones and the $B(M1)$ values are smaller, where the $B(M1)$ values
are small overall.  As seen in Fig.~\ref{fig:103rh}, the wobbling energies
increase with $I$ indicating longitudinal wobbling.  The Ru isotopes in
Fig.~\ref{fig:Ru} are similar to the Pd isotopes, which are discussed above.

From Figs.~\ref{fig:Pd}, \ref{fig:Rh} and~\ref{fig:Ru}, it can be seen that
large strength for the $B(M1)$ transition corresponds to small wobbling energy and
vice versa. Both quantities reflect the coupling strength of the quasiparticle to
the triaxial core. Strong coupling enhances the $B(M1)$ values and it favours
the transverse wobbling, which corresponds to small wobbling energy.  For
example, in the Rh chain QTR calculations the coupling depends on the $\gamma$
parameter of the TR core.  When $\gamma$ $\approx$ $30 ^{\circ}$, the wobbling
energy is not close to the experimental one, and the $B(M1)$ values of QTR
calculations are smaller than those of the QTD calculations. But when $\gamma$
is adjusted to reproduce the experimental wobbling energy, the $B(M1)$ values
are increased substantially, becoming larger than in the QTD calculations.

\section{Summary}
\label{sec:summary}

In this paper the triaxiality of odd-mass nuclei was explored by investigating
the coupling of an odd high-$j$ quasiparticle to the triaxial core represented
by the even-even neighboring nuclei.  Two kinds of core models were studied.
The \enquote{soft core} takes the rotation-vibrational motion of the shape into
account.  In this description, the core was described by a Bohr Hamiltonian with
a simple few-parameter potential, which provides soft confinement both in
$\beta$ and $\gamma$.  In particular, we adopted a triaxial Davidson
potential~\cite{collectiveH}, solved numerically using the Algebraic Collective
(ACM) approach.  The parameters were determined by fitting the energies of the
lowest collective states of the even-even neighbors.  The coupling of the odd
quasiparticle to this soft core was then described by means of the
quasiparticle-core coupling model of Ref.~\cite{cqpc}, resulting in a
description of the odd-mass nucleus which we termed the quasiparticle triaxial
Davidson (QTD) model.  The resulting energies and electomagnetic transition
probabilities were compared with those obtained by coupling the quasiparticle to
a \enquote{rigid core} described by the simple triaxial rotor model, in what we referred
to as the quasiparticle triaxial rotor (QTR) model. This represent the limiting
case of the QTD model where a stiff potential for the vibrational modes
leaves only the rotational motion active.

We studied the isotopic chains $^{101-115}$Pd, $^{101-115}$Rh, and
$^{107-113}$Ru.  These examples paint a general picture. The ``rigid core'' QTR
calculations gave for some cases the transverse wobbling pattern, where the
wobbling energy decreases with angular momentum then increases after some point,
but in most cases the longitudinal wobbling pattern, where the wobbling energy steadily
increases with angular momentum.  The competition between transverse and
longitudinal wobbling depends on the coupling strength between core and
quasiparticle, which is determined by the triaxiality of the core and the
position within the $j$-shell (top or bottom: transverse orientation; middle:
longitudinal orientation). Only a few of the QTR calculations resulted in a
transverse wobbling pattern, when the triaxiality parameter of the TR was close
to $30^\circ$ and the odd quasineuton was nearly a pure particle.  The QTD
calculations always gave the longitudinal wobbling pattern, even for the cases
where the QTR model provided transverse wobbling.  The reason is that in
contrast to the rigid TR core the soft TD core takes large excursions into
regions of small triaxiality.

Besides the different $I$ dependence of the energy difference, the soft core
leads to weaker $B(M1)$ values for the $\Delta I=1$ transitions from the
wobbling band to the yrast band.  The $B(E2)$ values for the same transitions
for both types of cores are collectively enhanced and not very different.

The tranverse wobbling pattern is observed experimentally. For the nuclei
studied here it appears in a volatile way for $^{101}$Pd, and much more clearly
for $^{135}$Pr \cite{pr135} and neighboring nuclei. In all cases, only the rigid
TR core is capable of accounting for the observed pattern, while the soft TD
core leads to longitudinal wobbling (see Ref.~\cite{thesisLi} for
$^{135}$Pr). This means that the coupling of the odd particle does not
sufficiently polarize the phenomenological TD core, such that the tranverse
wobbling pattern of the rigid core appears. 
The \enquote{soft core} derived from the neighboring even-even
nuclei coupled to a quasiparticle no longer describes the odd mass nucleus,
which is better described by a \enquote{rigid core} coupled to a quasiparticle.

The important implication is that the presence of the quasiparticle changes the
rigidity of the \enquote{core}. The CQPC model assumes that the collective
degrees of freedom of the core and the quasiparticle degrees of freedom are
independent. However, the collective excitations of the core are superpositions
of two-, four-, $\ldots$ quasiparticle excitations. The presence of the odd
quasiparticle blocks some of these quasiparticle excitations, which modifies the
response of the core relative to that which is observed in the even-even
neighbors.

These modifications of the core and the ensuing changes of the coupled system
have been studied for spherical nuclei in the framework of nuclear field
theory~\cite{NFT}.  The Klein-Kerman (KK) equation of motion
method~\cite{kerman1962:eom} accounts for these effects in principle. However,
the anti-symmetrization between quasiparticle and core has only be taken into
account for the triaxial rotor, which leads to the self-consistent triaxial
rotor model. The modifications of the rotor core solved the longstanding
Coriolis attenuation problem of the conventional particle rotor model (see
Ref.~\cite{Klein00} and references therein).  For more general core models the
KK approach became became a practical tool only after the introduction of the
D\"onau-Frauendorf approximation (DF), which neglects the core modifications
terms (giving the KKDF method).  The resulting CQPC formalism was used in this
paper (for a discussion of the DF approximation see Refs.~\cite{cqpc,Klein04}).

A promising new avenue is the triaxial projected shell model
(TPSM)~\cite{sheikh1999:tpsm}. The TSPM has been demonstrated~\cite{Jehangir21}
to be capable of describing the energies and quadrupole matrix elements of both
$\gamma$-rigid and $\gamma$-soft even-even nuclei, by diagonalizing a pairing
plus quadrupole-quadrupole Hamiltonian, in a basis of states comprised of
multiquasiparticle excitations in a fixed triaxial mean field, projected onto
good angular momentum.  
The TPSM also correctly describes transverse wobbling in $^{135}$Pr~\cite{Sensharma19}. 
It is of interest to study neighbor $^{134}$Ce within the framework of the TPSM in order 
to exemplify the  
core modification as a consequence of the exclusion principle for fermions.
Work along these lines is in progress.

\begin{acknowledgments}
This material is based upon work supported by the U.S.~Department of Energy,
Office of Science, under Award No.~DE-FG02-95ER40934.
\end{acknowledgments}

\end{document}